# A Game-Theoretic Frame Work for Studying Dynamics of Multi Decision-maker Systems

**Mohammad Rasouli**

Department of Electrical Engineering, Sharif University of Technology

NO 23, Yas Building, 26th St, Velenjak St, Tehran, Iran

Cell phone: +98-912-2992078

mrasouli@ee.sharif.ir

*Abstract-* *System Dynamics' (SD) main aim is to study dynamic behavior of systems based on causal relations. The other purpose of the science is to design policies, both in initial values and causal relation, to change system behavior as we desire. Especially we are interested in making system's behavior a convergent one. Although now SD is mainly used in situations of single policy maker, there are major parts of situations in which there are multi policy makers playing role. Game Theory (GT) is an appropriate tool for studying such cases. GT is the theory of studying multi decision-maker conditions. In this paper we will introduce GT and explain how to apply it in SD. Also we will provide some examples of microeconomic systems and show how to use GT for studying and simulating dynamics of these example systems. We will also have a short discuss on how SD can help GT studies.*

**Introduction**

System dynamics (SD) is a powerful tool for studying dynamics of systems. It can be used for studying a great variety of systems including social, economic, political systems and etc. What does SD helps us to know? We can list them as following:

1) SD helps us to simulate a system from its initial state (state at present time) and see what will happen next in every step of time.
2) As SD helps us to foresee future of the system, it can help us to see effect of every policy we choose at present. So it helps us in policy making and choosing policies to have our system in desires state in future.

What do we need to use SD? We need to identify causal relations among different parts of the system and apply appropriate equations for these relations if we want to have quantitative simulations. Note that most of the time we desire system's long term behavior to be a steady state situation. It means we like our systems to be stable ones and not oscillating or divergent ones. Now, in most of system dynamics applications we consider a single policy maker situation. It means that we are the only decision maker of the systems and after our decision others will be forced to do certain actions because of causal relations. But this is not the real situation in a large number of systems.

Game theory (GT) is a common tool in studying distributed systems. GT has a widespread application in studying systems, like economic, social, engineering and etc. What does GT helps us to know? Here are list of different purpose of applying game theory:

1) Game theory helps us to simulate system with independent decision makers. it helps us to see what the outputs of the games are.
2) Game theory is widely used in designing games (Systems) in which although different players take their own action independently but output of the game (system) is just as we will. This application of GT is called "Mechanism Design".

What do we need to use GT? Game theories main assumption is *rationality* of the players. Here rationality means that every player just acts as to maximize his own benefit. Also players are aware of rules of the game and available actions of every other players of the game and they use logical deductions to choose their own actions. With these basic assumptions we need to define players, actions and payoff of every outcome of the game for every player to use GT. Game theory is mainly focused on long term behavior of the systems, also focused on finding convergent possible outputs of every game. Although game theory is strong in modeling long time behavior of systems it tells a few about transient time of the games. Also whether game with a special initial condition converge to their convergent output or not, and if yes to which of possible equilibrium output they converge.

It can be seen from above discussion that game theory and system dynamics have same purposes and same applications, but each one has some strength point and some weaknesses. In a general conclusion we can say SD studies dynamic and centralized systems while GT studies static and distributed systems. Mastering in both tools and applying both in studying systems can help in better understanding of system behavior and make more progress.

In this paper we will shortly introduce game theory we will discuss on how to apply it in system dynamics. We will provide some examples. Examples are chosen from micro economy, because both game theory and system dynamics are widely applied in this field. We have assumed that reader is well familiar with SD. So we do not explain it. At the end we will mention how SD can be used in game theoretic frame work studies.

**Brief Introduction to Game theory**

Game theory is the theory of multi decision maker situations, where every decision has direct effect on others preferences and benefits. Game theory has been developed to model different kind of games. In a general division we can divide games in four groups: 1) strategic games 2) extensive games with perfect information 3) extensive games with imperfect information 4) cooperative games

The base for this division is whether games are with cooperative actions or not, whether they are simultaneous or in consequence and whether players have perfect information or not. Perfect information is that every player knows others possible actions and their preferences on the outcomes of the game.

In this division the first kind of game, strategic games, are defined as simultaneous, non cooperative and with perfect information games. These assumptions make these games easier to model. These games are applied in wide area of applications. We will continue our discussion on this kind of games. These games consist of 3 elements:

1) a set of players : N={1,2,3,…,n}
2) a set of actions for each user Ai, for every i a member of N
3) a utility function for every user, $u_i : A \rightarrow R$, which defines his preference for every outcome possible outcome of the game, which are members of $A = \times_{i \in I} A_i$

When the game is played, each player i selects an action from his own set of actions Ai. These selections are made without any knowledge of the selections made by others.( This is the simultaneous assumption of the game) The selections of all players taken together define an action profile, $a \in A$, and each player i receives the payoff ui(a). Rationality assumption means each player wants to maximize his own payoff. Ordinarily, we assume that a player is not limited to choosing actions directly from Ai. Instead, players are allowed to choose "mix strategies" or mixed actions which are probability distributions over Ai. But here we do not continue in mix strategies. As their simulation is just like pure strategies we limit this paper in pure strategies.

Once such a game has been defined, game theory defines a solution concept which attempts to specify what we should "expect" to occur if rational players play the game. The most widely known solution concept is the Nash Equilibrium. For convenience, we will sometimes write an action profile $a \in A$ as $(a_i, a_{-i})$ where ai denotes the action chosen by player i and $a_{-i}$ denotes the actions chosen by everyone else. An action profile $a \in A$ is said to be a Nash Equilibrium if for every player $i \in I$, and for every a_i from his action set Ai, $u_i(a_i, a_{-i}) \geq u_i(a\_i, a_{-i})$.

That is, an action profile is a Nash Equilibrium if no player can gain by unilaterally deviating from the specified profile. When players play rationally, they will have Nash point as the output of the game if the game just has one Nash equilibrium point. If there are more than one Nash points they will converge to one of them as the game is played again and again unless the game will not converge to any point. Initial actions of players define to which Nash point the game will converge. John Nash proved that if mixed strategies are allowed, then at least one equilibrium exists for every finite game (A finite game is a game with finite sets I and Ai.) There is an important of interpretation of Nash equilibrium and whether it is a steady state outcome of a system or not. We are not going to open all respects of the subject but we will introduce iterative interpretation of Nash equilibrium: in a simultaneous game (strategic one) if we repeat the game such that every player does maximize his instant payoff and there is no strategic link in different repetitions of game (players can choose every action independent of their previous actions), then the game outcome will converge to its Nash equilibrium points or it will diverge.

To find Nash points in a game we need to introduce Best Response functions.

For any a-i action profile chosen by other players we define $BR_i(a_{-i})$ to be the set of player i's best actions given a-I, meaning :

$BR_i(a_{-i}) = \{a^*_i \in A_i: u_i(a_{-i}, a^*_i) \geq u_i(a_{-i}, a_i) \text{ for all } a_i \in A_i\}$

Now it can be easily seen that a Nash equilibrium is a profile a* of actions for which

$a^*_i \in BR_i(a^*_{-i})$ for all $i \in N$

what does these mathematic equation tell us is that every player from the assumption of rationality, does one of his best response actions to what he thinks the others choose as their action profile. Now in Nash point every player's action is a best response to others. So no one could do anything better to increase his payoff. For finding Nash points we have to determine every players best response and then look for the point where these best response functions ( or mappings) meet each other. Now we are ready to apply GT in SD.

**Applying Game theory in System Dynamics**

Assume the situation in which you are an economic supplier that produces a certain good. There are some other suppliers in your market that also product a certain good. Now each of you have to decide quantity of your product and the price in the market is set from the total products supplied in the market. There is no cooperation among suppliers and each one wants to maximize his own payoff. The question is that how much you should set your own product?

It can be seen that here there is a multi decision making system. You can also set your own policy but you have no power on others'. How can you simulate this system and find your best policy?

Let's start from a monopoly example. Consider a monopoly market ( market that has only one supplier.) supplier's quantity of product is shown with q. the price in the market ( with the assumption of market clearance and equilibrium of supply and demand) is in relation with q: p=(1000-q)/10. Your cost of product is a function of your product, C= -q^2+100*q. the benefit of the supplier is:

Benefit= p*q-C(q)

From maximizing your benefit you can find quantity as a function of price.

For simplicity of model we assume that supplier supplies his entire stack to the market and does not accumulate goods there. Here are the causal loop and the model.[1] The model is attached in supporting materials.[2] There is negative loop between price and quantity that makes a converging behavior.

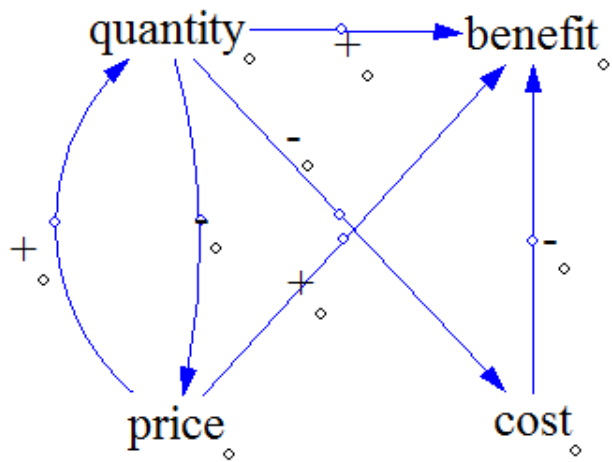

---

[1] All models and simulations are done by Vensim software.

[2] It is recommended to see models and its parameters' formulations for a better understanding. Please contact the author for access to the supporting files.

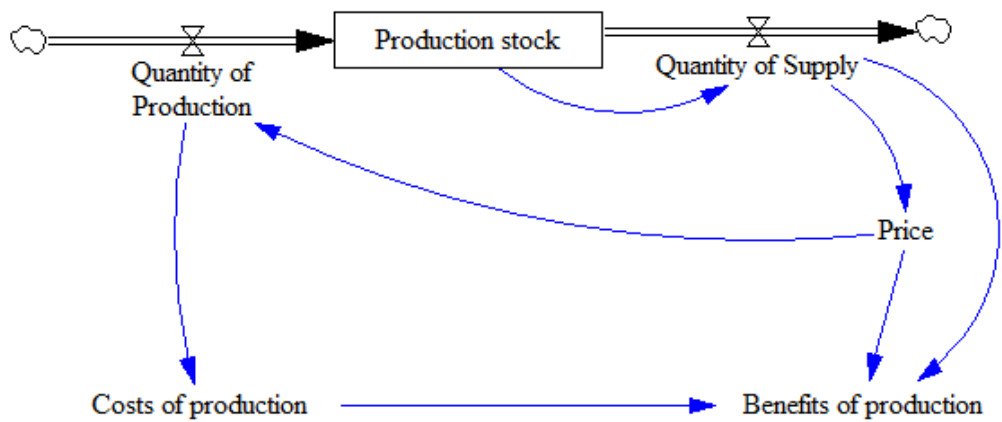

Figure 1: a) causal model for monopoly b) Vensim model for monopoly

Here we have used quantity to update price and price to update quantity in every step. Here are the results.

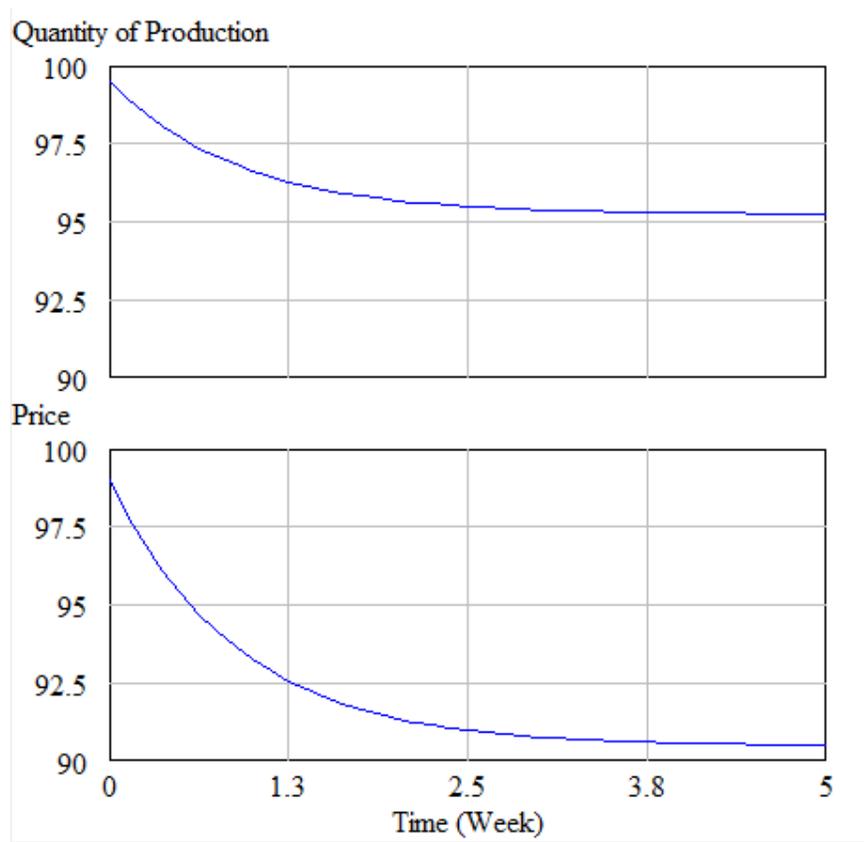

Figure 2: monopoly simulation results.

Now let's go back to the first question. If consider a 3 supplier market each with a producing qi quantity of a certain good. Each one has a cost of Ci(qi)=-qi^2+100*q. and the price in the market is set as: p=(900-Q)/10 where Q is total product q1+q2+q3. For simplicity of the model assume suppliers supply their entire stack to the market. Here rises the problem. We, as supplier number 1, can only set our production quantity, q1. But if we want to model the market we must somehow model others behavior. Let's use Game theory and look at the problem again. Here we have a strategic game. The players are the 3 supplier. Their set of actions is to choose a number larger than or equal tob0: qi>=0. Their payoff for output (q1, q2, q3) is:

Payoff i = qi*p – C(qi)= qi*(900 – q1 – q2 – q3)/10 - (- qi^2+100*qi )

Suppose that suppliers decide their supplying quantity at the beginning of every week. So we can say that suppliers play a strategic game at the beginning of every week. We have simulates the situation like a game. Now GT acclaims that if we consider the player to be rational and play rationally, then every player has to choose his best response to other players. Let's find the best response function. According to definition we know that best response function is the choice that maximizes the payoff. By differentiating the payoff 1 function by q1 variable we have:

BR1(Q-1)= q1=(10 + (q2+q3)/10) / 1.8

From symmetry we can say the 2 other best response function is the same. So at the beginning go every week suppliers try to make a belief (guess) about others strategy and take their own as the best response. This is just the iterative interpretation for strategic games.

Now we can model the game and we are not worry any more of the others policy or strategy. We can place their best response as their action (policy).

Now we want to simulate the situation. Here is the causal loop and Vensim model[3]:

---

[3] We have set different initial value for suppliers.

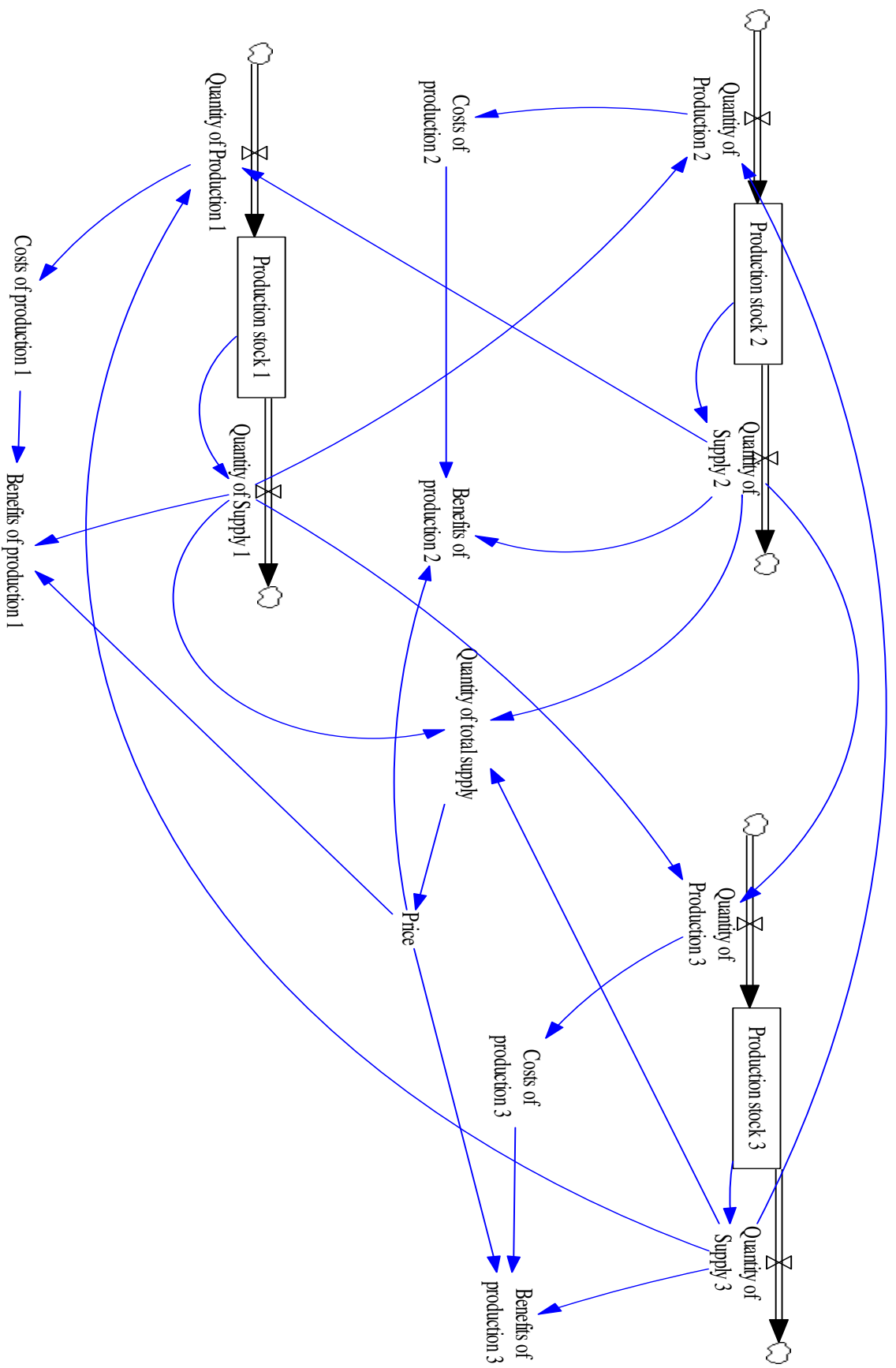

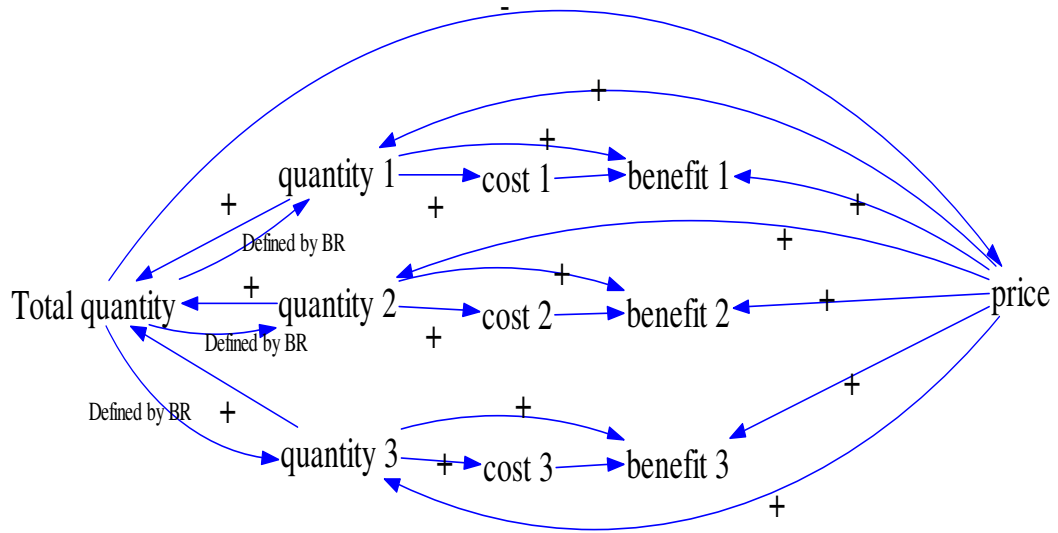

**Figure 3: multi supplier market a) causal loops b) Vensim model**

We can see the negative loops among total quantity, price and each supplier's quantity which makes a converging behavior.

Here are the results:

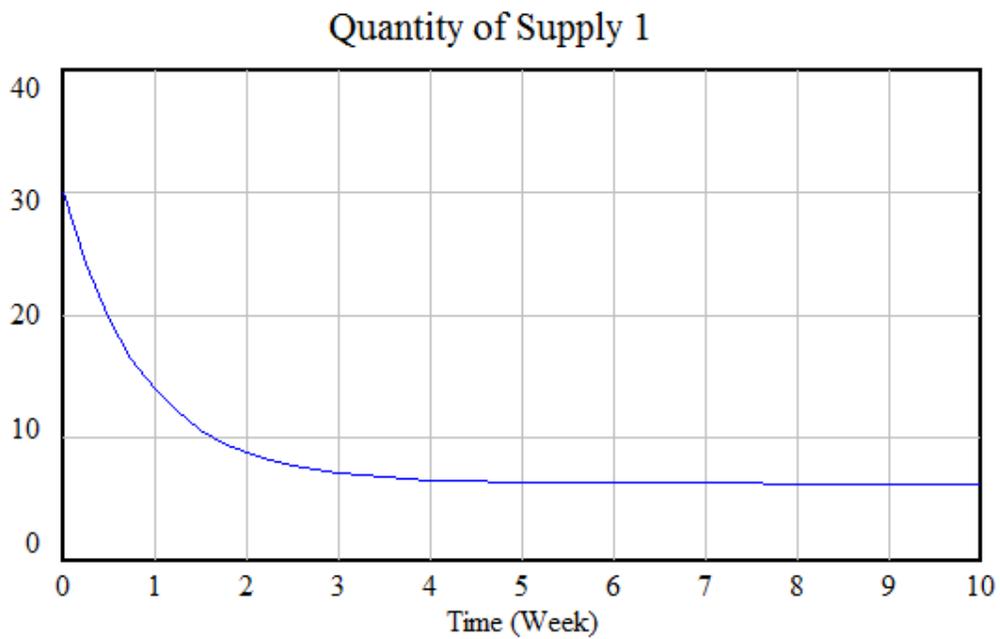

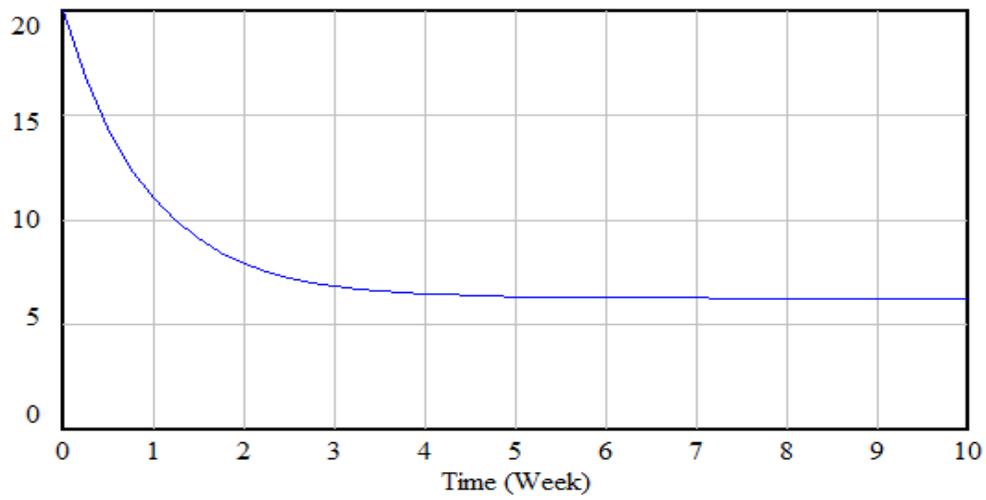

Quantity of Supply 2

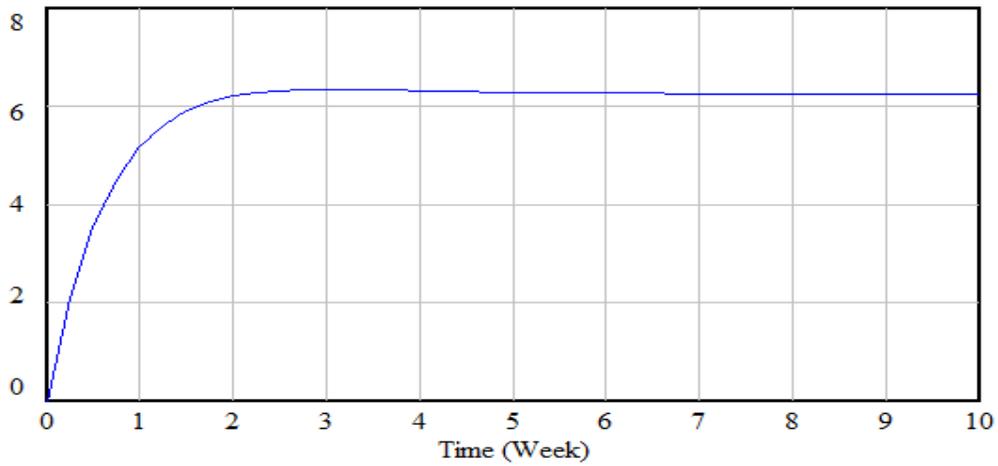

Quantity of Supply 3

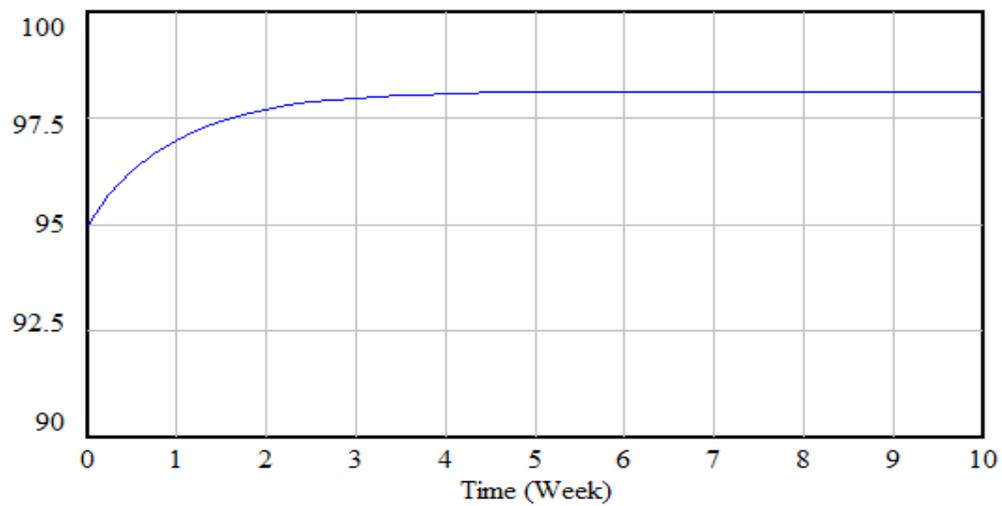

Price

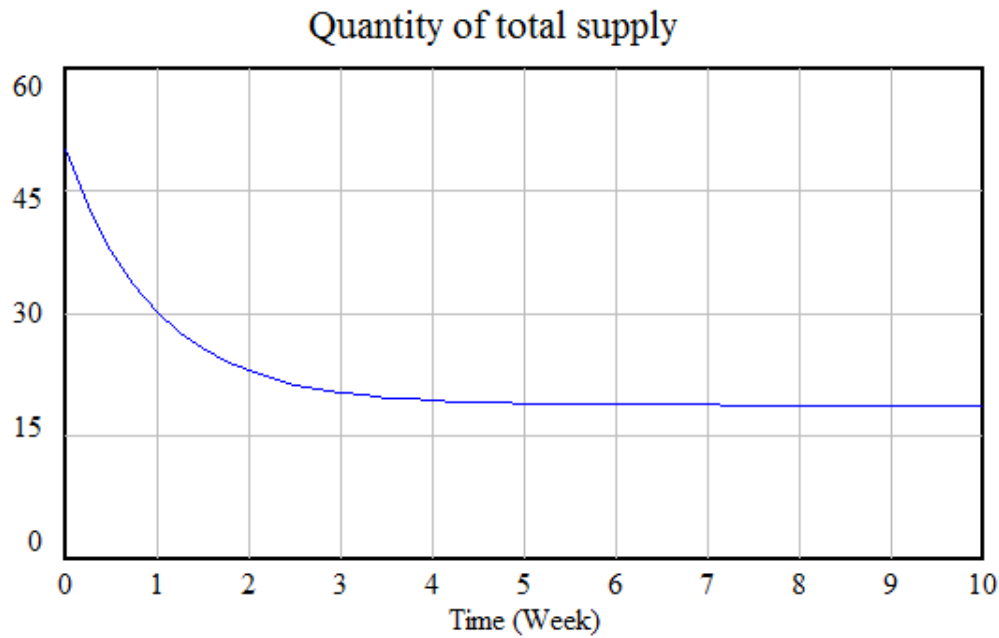

**Figure 4: multi supplier market simulation results: a) quantity of production of players 1,2 and 3 b) price c) total product**

What did we expected to take place? From game theory We expect that the game converges to one of its Nash Equilibrium point (if it converges). By finding NE points , the point where BR functions meet each other we can see that this game has only one NE point and it is: (q1*,q2*, q3*) = (6.25, 6.25, 6.25) and because there is only one NE point the game *must* converge to that. It is completely what we have reached from simulations.

Now let's consider another example that has more than one Nash equilibrium point.

Consider another market again with 3 suppliers. The production procedure of the good generates pollution which is dangerous for the environment. So the government will increase the tax level if the total pollution is more than a certain limit to control the pollution. Assume that volume of pollution generated is in a linear relation with quantity of production. The price of good is100 and for simplicity of model we assume that the only cost for the supplier is the tax. When the total production is less than 200 units the tax per unit is 50 otherwise it is 150. Again the problem is that we are producer number 1 and have to make policy on our behavior, but there are others who make decisions separately. We try to look at the system like a game: players are the 3 supplier. Their action set is to choose qi>=0.their payoff is:

Payoff i=qi*price-qi*tax

Tax=if (total product>200) then 50 else 150

The Best Response function for player 1 is:

BR1 (q1) =min ((200-q1-q2), 0)

From symmetry the others BR functions are same. Now we use iterative interpretation of game theory and consider that suppliers choose their amount of production at the beginning of every week simultaneously. So they play a strategic game every week.

Now we can model the game. Here is the Vensim model:

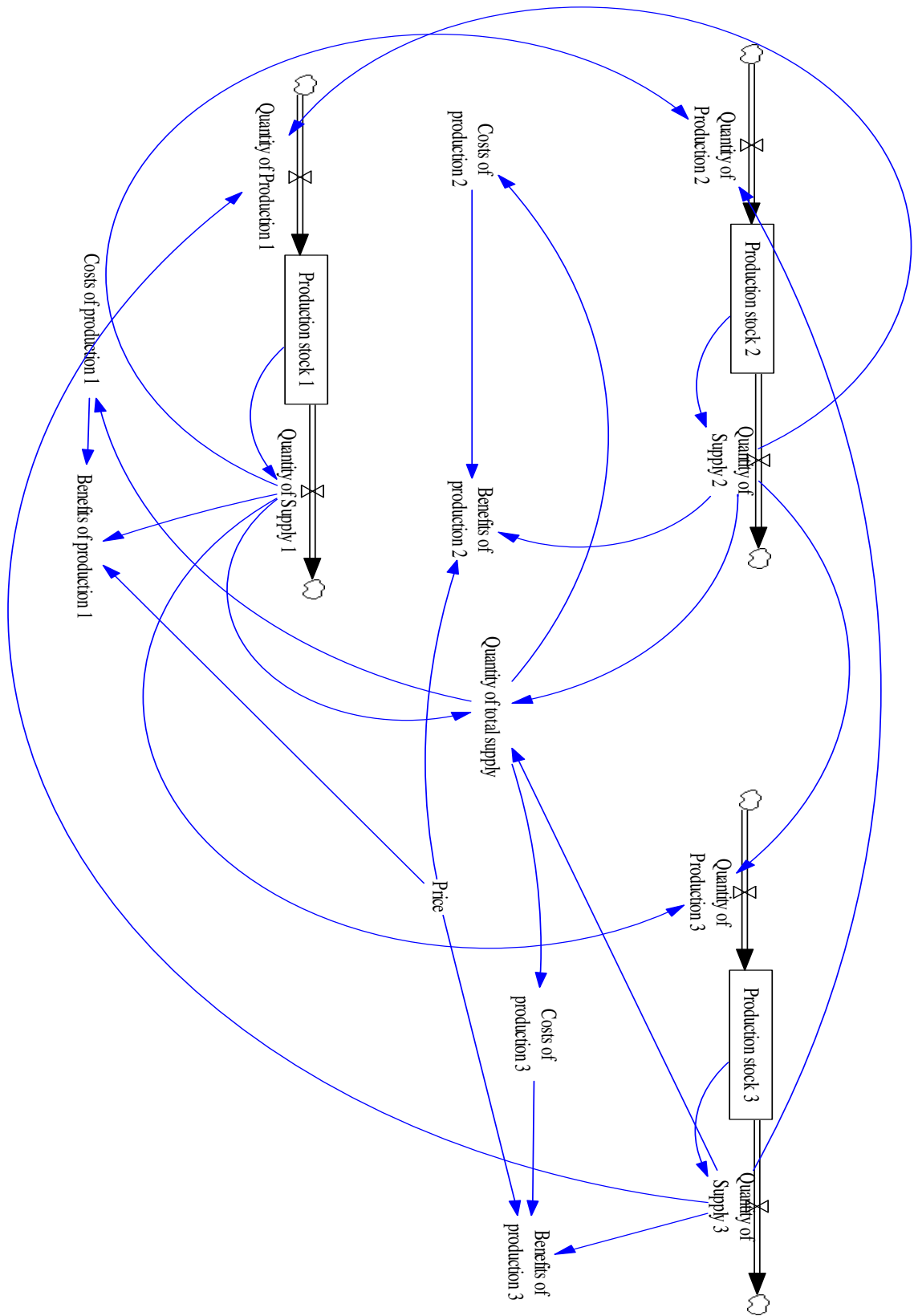

**Figure 5: Vensim model for polluting suppliers**

Here are the results for two different initial value simulations. First we have set (q1, q2, q3) = (10,100, 50). In second we have set (q1, q2, q3) = (300,100, 50).

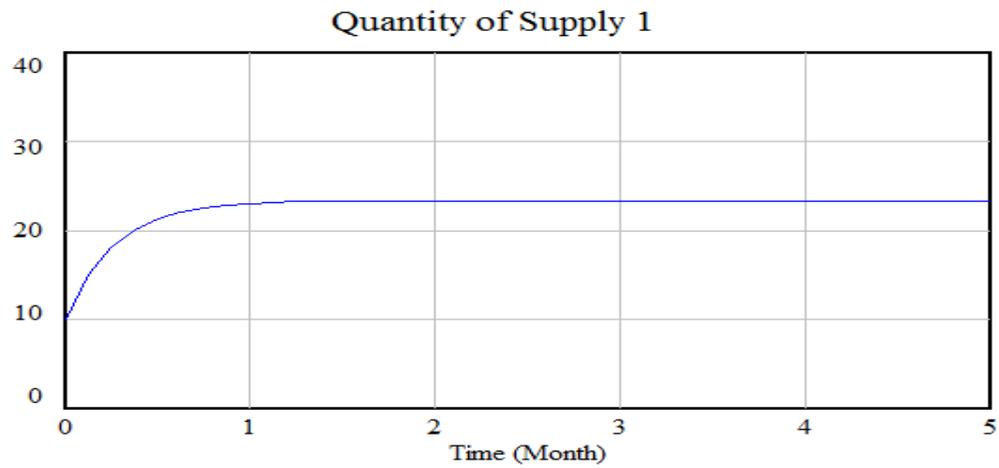

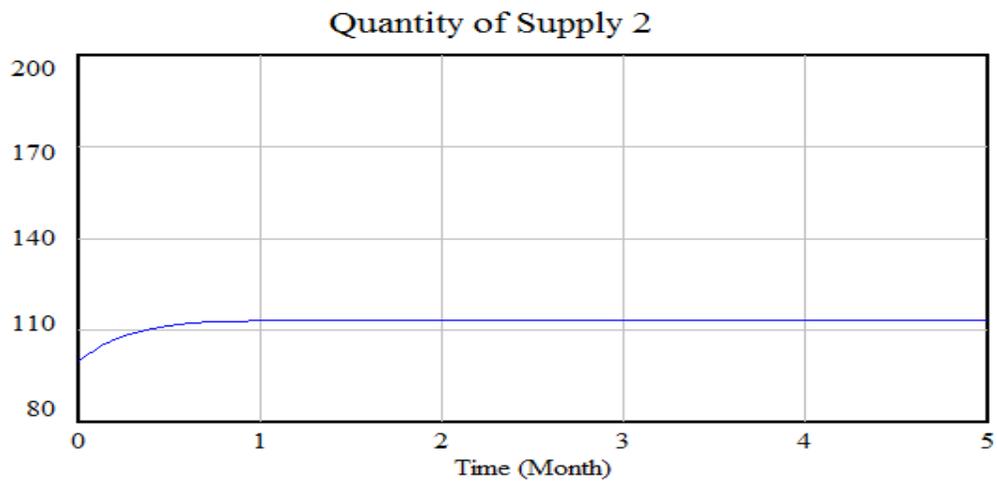

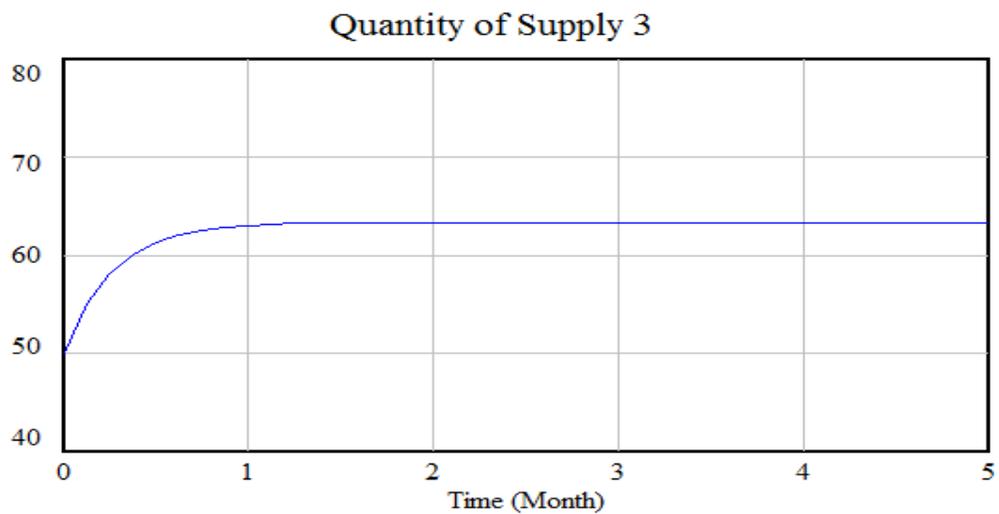

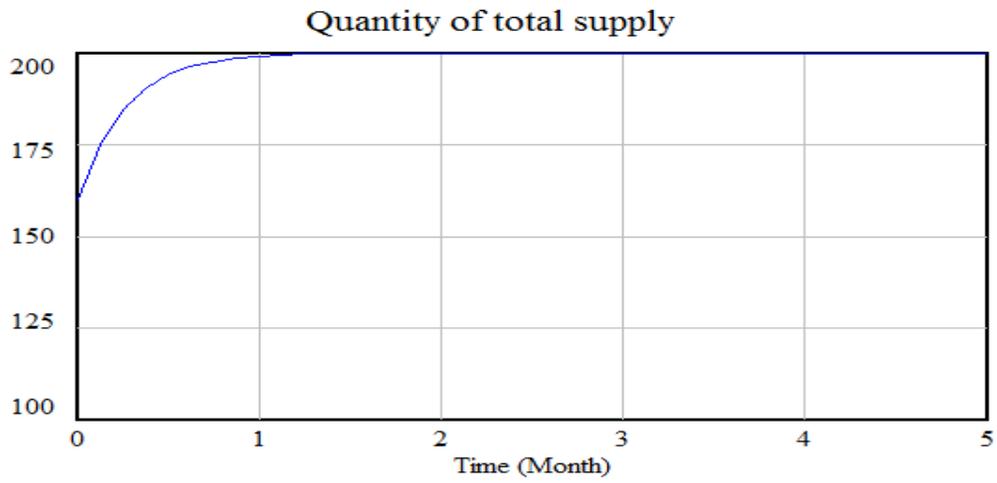

**Figure 6: results for polluting suppliers with initial set 1**

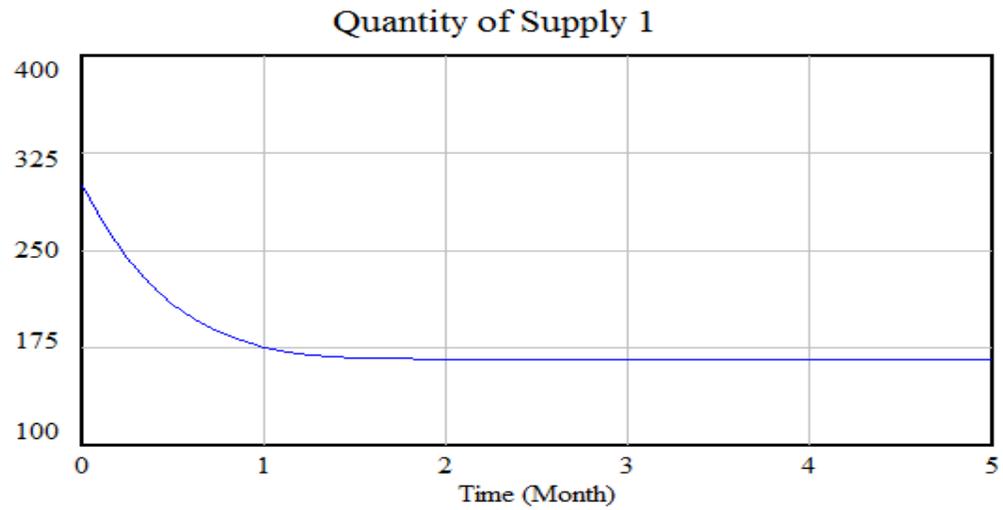

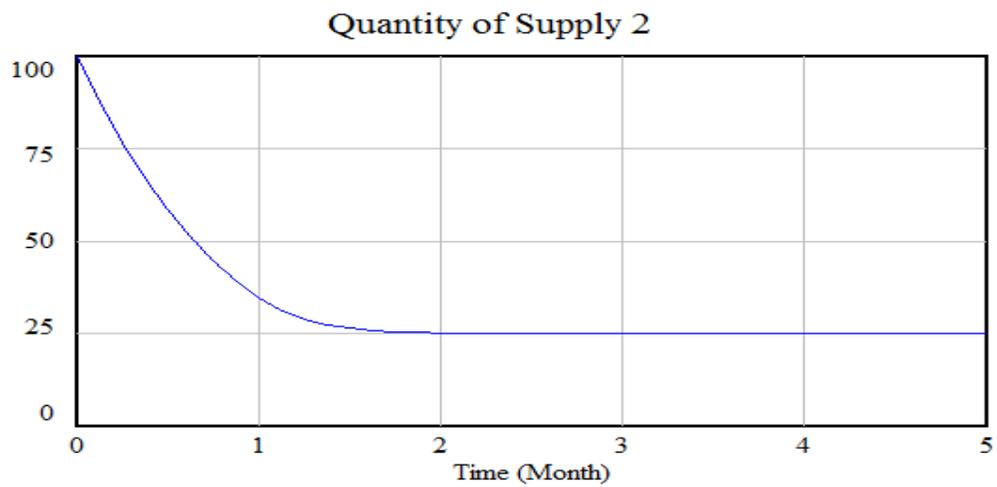

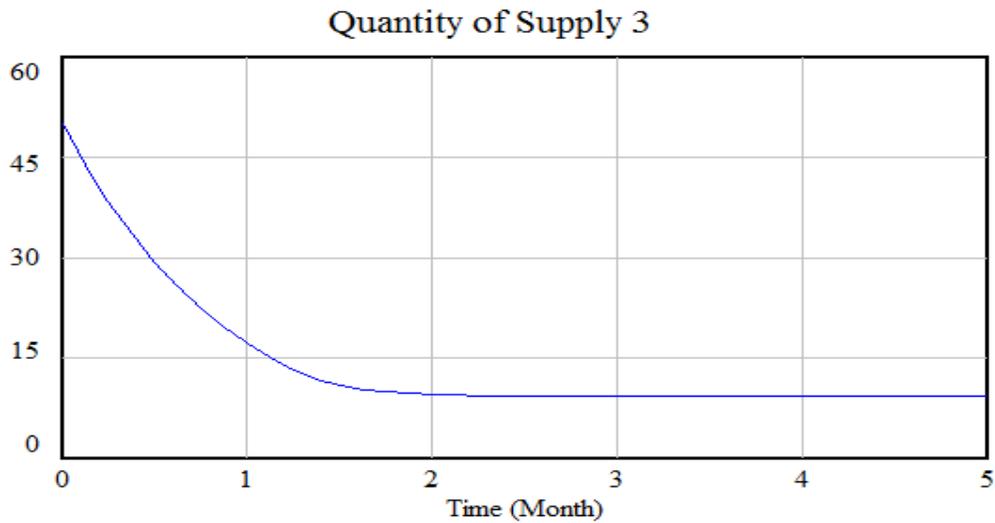

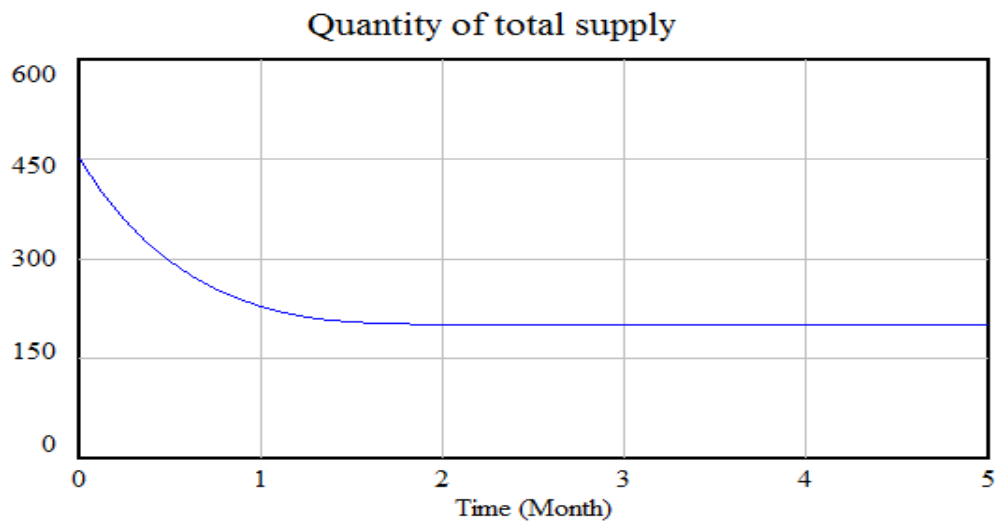

**Figure 7: results for polluting suppliers with initial set 1**

What are the Nash equilibriums of the game?

Every set (q1, q2, q3) that all qi>=0 and q1+q2+q3=200 is a Nash equilibrium for this game, because in these points BR functions meet each other. We see that in above examples the total product is always 200 in steady state, this acknowledges our belief. But which Nash equilibrium will be our result? This depends on initial value. Here is where GT does not have any response, we will discuss this question more in next part.

Now that we have seen some examples we provide a total method for simulating multi decision maker system.

First step is to simulate the system as a game. It means to define players and action profiles and payoffs. Note that a system is a game if every player's action does have effect on others' payoffs. Also note that we have discussed strategic games (simultaneous perfect information games) in this paper. If you the decision are not simultaneous or somehow sequential, or there is no perfect information situation you should use other kind of games. For simultaneous game you should divide time to discrete times and assume every player does choose his action at the beginning of a time section and cannot change it until it finishes. Also at the beginning of the time part every player will assume others will continue their own action for next part (making a belief about others actions). Now every player chooses his own new strategy as the best response function to his belief. This is the golden point game theory helps us in simulating dynamic systems. In fact we have a one step delay system in which all players' decisions together makes their next step decision.

Your policy here is your first (initial) action. And by setting your policy and making a belief about others' first action you can rub the simulation and see the result.

Game theory has more to help SD. It helps us to find all steady state situation of a system. There may be mistakes in generalizing a steady state of a system for a special initial value to other initial values. Also it helps us to see whether our system has a steady state or not, and if yes is our willing outcome a steady state or not, before we spend time on finding policies to reach our desired steady state.

**Using system dynamics in game theory approaches**

There are ways in which SD can complete GT. A main problem in GT is to find NE points. In fact it is difficult in a lot of our real system. There are some theories on existence of NE points for a game. So in systems where we now there are NE points we can use SD to find those points. Note that we now an iterative game as discussed in iterative interpretation of games will converge to an NE point (if it does converge). SD can be even used for testing if there is any Nash point. By running a system with iterative steps, in case of convergence of actions we can result that there is a Nash and also set the convergence point as NE point of the game.

But SD has more to help GT than this. Let's return to our polluting suppliers' example. The main problem was the NE point we would converge to with a certain initial values. It can be generalized to if we converge to NE points or we will diverge (infinite amount of production in our examples) or oscillate with an initial value set.

Here game theory has no answer and other approaches should be used. SD presents a satisfying response by simulating the system step by step. Sometimes we have limits in transient time states of a system. Such limits are mostly in engineering systems. Here again game theory does not have any guaranty, where as SD shows us complete transient time of a system.

**Deduction and future research comments**

In this paper we discussed how game theory with its abilities to simuate steady state in multi decision making systems can help system dynamics. We tried to show this via some examples. In fact both game theory and system dynamics are utilities necessary for one who wants to study systems. For future works, studying in other kind of games and their application in system dynamics are recommended. Also there is wide area of applying system dynamics in game theory approaches.

**Mohammad Rasouli** was born in Tehran, Iran, on September 21, 1987. He received B.S. degree in electrical engineering from Sharif University of Technology, Tehran, Iran in 2008 (with Honor third rank) . He is currently working toward M.S degree in Sharif University of Technology (SUT), Tehran, Iran. Since summer of 2007, he has been working with Dr.Mashayekhi and passed some courses such 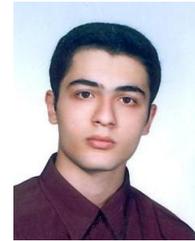 as System Dynamics I, and some courses with Dr.Nilli such an Introduction to economy and game theory. He is working on game theory in networks. He also has an experience in managing some Projects, and has been head of Sharif University of technology students' congress in 2009.